\documentclass{article}
\usepackage{spconf}

\usepackage{amsmath, bbm}
\usepackage{amssymb}
\usepackage{amsfonts}
\usepackage{amsthm}
\usepackage{algorithm}
\usepackage{algorithmic}
\usepackage{nicefrac}
\usepackage{bm}
\usepackage{hyperref}
\usepackage{cite}
\usepackage[utf8]{inputenc} 
\usepackage[T1]{fontenc}
\usepackage{graphicx}
\usepackage[font=small]{caption}
\usepackage{tikz}
\usetikzlibrary{positioning}
\usepackage{pgfplots}
\usepackage{epstopdf}
\usepackage{subcaption}
\usepackage{multirow}
\usepackage{lipsum}
\usepackage{diagbox}
\usepackage{soul}

\colorlet{bl_holder}{green!10!orange!90!}

\newtheorem{Problem}{Problem}

\input{mysymbol.sty}

\title{Deep Demixing: Reconstructing the Evolution of Epidemics using Graph Neural Networks}

\name{Gojko Čutura$^\star$, Boning Li$^\dag$, Ananthram Swami$^\ddag$, and Santiago Segarra$^\dag$
\thanks{Research was sponsored by the Army Research Office and was accomplished under Cooperative Agreement Number W911NF-19-2-0269. 
		The views and conclusions contained in this document are those of the authors and should not be interpreted as representing the official policies, either expressed or implied, of the Army Research Office or the U.S. Government. 
		The U.S. Government is authorized to reproduce and distribute reprints for Government purposes notwithstanding any copyright notation herein.
		\newline
		Emails:  cg203066m@student.etf.bg.ac.rs, \{boning.li, segarra\}@rice.edu, ananthram.swami.civ@mail.mil.}}
\address{$^\star$University of Belgrade, Serbia \hspace{1cm} $^\dag$Rice University, USA  \hspace{1cm}  $^\ddag$US Army Research Lab., USA}

\begin{document}
\ninept
\renewcommand{\baselinestretch}{0.95}
\maketitle
\begin{abstract}

We study the temporal reconstruction of epidemics evolving over networks. 
Given partial or aggregated temporal information of the epidemic, our goal is to estimate the complete evolution of the spread leveraging the topology of the network but being agnostic to the precise epidemic model.
We overcome this lack of model awareness through a data-driven solution to the inverse problem at hand.
In particular, we propose DDmix, a graph conditional variational autoencoder that can be trained from past epidemic spreads and whose latent space seeks to capture key aspects of the underlying (unknown) spreading dynamics. 
We illustrate the accuracy and generalizability of DDmix and compare it with non-graph-aware learning algorithms through numerical experiments on epidemic spreads simulated on synthetic and real-world networks.

\end{abstract}
\section{Introduction}\label{sec:intro}
\red{
}

Networks or graphs have emerged as useful models to represent complex interconnected systems and data defined on them~\cite{jackson2010}.
These representations have wide applicability across multiple domains, including neuroscience~\cite{Medaglia2017brain}, sociology~\cite{newman2002random}, and urban planning~\cite{dogrusoz2007modeling}. 
In an attempt to better understand and learn from data defined on networks, classical signal processing and machine learning methods have recently been extended to encompass this data type~\cite{ortega2018graph}.
These novel tools have shown remarkable performance in a variety of popular network science problems such as node classification~\cite{kipf2016semi_supervised}, link prediction~\cite{zhang2018link}, and several inference tasks  related to partially observed network processes~\cite{segarra2017blind, zhu2020estimating}.

Within the range of network processes, the use of graphs for modeling and understanding \emph{epidemics} is a fertile subfield~\cite{kiss2017mathematics}.
Networks provide versatile modelling tools where nodes might represent anything from single individuals~\cite{zheng2019node_based} to large cities~\cite{pujari2020multi_city} and edges can encode different mechanisms of disease propagation -- airborne, contact, or vector transmission -- across the nodes of the network.
\emph{Classical} graph features such as node centralities and connectivity measures can then be used to inform public health measures, e.g., immunization strategies and lockdown procedures~\cite{cliff2019network, block2020social}.

In this paper, we go beyond established classical methods and propose a novel neural network architecture to solve the challenging problem of temporally reconstructing an epidemic. 
More precisely, given partial or aggregated temporal information of the epidemic, we want to infer the evolution of the spread.
Being an inherently ill-posed inverse problem, one might rely on precise knowledge of the network process or structural features of the initial condition to solve this underdeterminacy~\cite{segarra2017blind}.
By contrast, we propose a \emph{model-inspired data-driven} solution.
Namely, we put forth a conditional variational auto-encoder (CVAE)~\cite{sohn2015learning} based on graph neural networks (GNNs)~\cite{kipf2016semi_supervised,gama2019convolutionalgraphs} that is trained to solve the inverse problem from available data and, importantly, whose architecture and training loss are inspired by the locality of the spreading mechanism.

\vspace{0.5mm}
\noindent{\bf Related work.}
We can model an epidemic spread as the temporal evolution of a signal defined on the nodes of a graph.
From this viewpoint, epidemic reconstruction boils down to different well-studied problems depending on the observation model.
More precisely, if the state of some nodes is observed and we want to infer the state of the remaining nodes, the problem can be modeled as the interpolation of graph signals~\cite{chen_2015_sampling, marques_2016_sampling}; if the final state of every node is observed, the problem resembles blind deconvolution on graphs~\cite{segarra2017blind}; and if a temporally aggregated signal for each node is observed, then the problem boils down to \emph{demixing} of graph signals~\cite{iglesias2018demixing}.
This last body of work is the one that best fits our observation model. 
However, existing tools~\cite{iglesias2018demixing} were derived for \emph{linear} network processes that cannot accurately model epidemic spreads, thus prompting the need for \emph{non-linear} methods as DDmix, the one derived here.

From the perspective of computational epidemiology, temporal reconstruction has also received attention~\cite{jombart2014bayesian,rozenshtein2016reconstructing}. For example, under the assumption that an epidemic spread follows the SI model, \cite{prakash2012spotting} recovers multiple source nodes from a single snapshot of the complete graph. 
Similarly, \cite{lappas2010finding} focuses on identifying key spreaders under the independent-cascade model.
In general, this body of work largely relies on the precise knowledge of the epidemic model or the time of contagion for a subset of nodes.
We depart from this paradigm and, instead, rely on observed past data to learn how to solve the reconstruction problem.

Related to our proposed architecture, CVAEs~\cite{sohn2015learning}
have been shown to be effective in solving prediction and data generation problems in non-graph settings~\cite{lim2018molecular, pol2019anomaly} and have been used for the reconstruction of videos from temporally aggregated data~\cite{balakrishnan2019visual}.
Motivated by this success, some of these tools have been extended to graph settings~\cite{kipf2016variational, gao2019graph} and mostly applied as node embedding procedures~\cite{hamilton2017inductive}. 
To the best of our knowledge, this is the first implementation of a graph CVAE for the study of epidemics and, generally, for the solution of inverse problems related to network processes.

\vspace{0.5mm}
\noindent{\bf Contribution.} 
The contributions of our paper are twofold: \\
i) We propose DDmix, a novel graph CVAE architecture for the temporal reconstruction of partially-observed network processes; and \\
ii) We successfully implement DDmix to infer the evolution of epidemics surpassing non-graph-aware deep learning methods in terms of accuracy and generalizability.

\section{System Model and Problem Statement}
\label{sec:problem}

We model our inter-connected system as an undirected graph $\mathcal{G} = (\ccalV, \ccalE)$ with $N$ nodes in $\ccalV$ representing individuals and edges $(i,j) \in \ccalE$ encoding the possibility of contagion between nodes $i$ and $j$.
The graph structure can be represented using the symmetric adjacency matrix $\bbA \in \reals^{N \times N}$, where $A_{ij} = A_{ji} = 1$ for all $(i,j) \in \mathcal{E}$, and $A_{ij} = A_{ji} = 0$ otherwise.
We model the state of the nodes at any given time using graph signals, i.e., maps $x: \ccalV \to \reals$ from the node set into the reals. 
Graph signals can be conveniently represented as vectors $\bbx \in \reals^N$, where the entry $x_i$ collects the value of the graph signal at node $i$.
Specifically, we consider a time series of graph signals $\{\bby^{(t)}\}_{t=1}^T$ where $\bby^{(t)} \in \{0,1\}^N$ and $y^{(t)}_i=1$ indicates that node $i$ was in the infected state at time instant $t$ while $y^{(t)}_i=0$ indicates that it was not infected at that time.

Although not assumed to be known by our proposed solution, for simplicity in this paper we focus on a particular epidemic model, namely the well-established SIRS model~\cite{zheng2019node_based}.
This is a parametric stochastic model that, given $\bby^{(1)}$, determines the probability distribution of the signals $\bby^{(t)}$ for $t>1$.
More precisely, the SIRS model is defined by three parameters -- the infection probability $\beta$, the healing probability $\delta$, and the probability of losing immunity $\gamma$ -- and every node is at one of three states -- susceptible (S), infected (I), or recovered (R).
At every discrete time step $t$: i) Susceptible nodes can get infected independently with probability $\beta$ by any of their infected neighbors in $\ccalG$, ii) Infected nodes recover with probability $\delta$, and iii) Recovered nodes become susceptible with probability $\gamma$.

\begin{figure*}[t]
	\centering
    \includegraphics[width=0.9\linewidth]{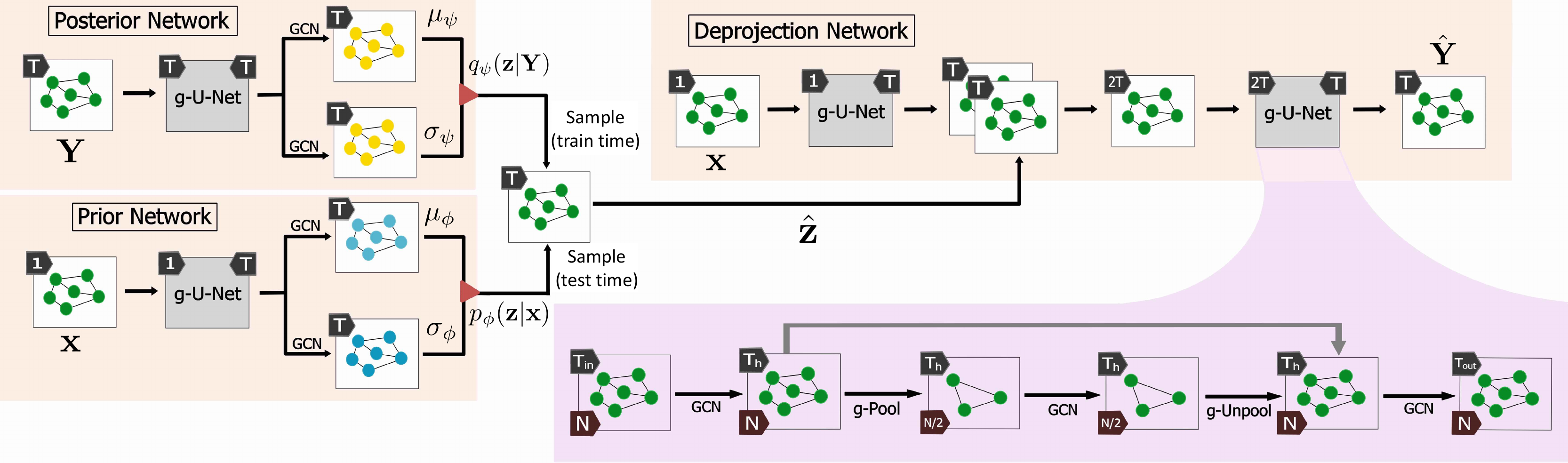}
    \vspace{-2mm}
	\caption{Overall view of our proposed deep demixing (DDmix) architecture based on graph neural networks for the temporal reconstruction of epidemics. The three subnetworks -- prior, posterior, and deprojection -- parametrize key functions in the assumed probabilistic model.}\label{fig:architecture}
	\vspace{-2mm}
\end{figure*}

Under an (unknown) SIRS spread $\{\bby^{(t)}\}_{t=1}^T$ of interest, our partial information is given by a single graph signal $\bbx = f(\{\bby^{(t)}\}_{t=1}^T)$ that captures some observable feature of the epidemic.
In this paper we focus on the case where $\bbx$ represents a temporal aggregation of the infections, i.e.,
\begin{equation}\label{E:aggrregation}
    \bbx = \frac{1}{T} \sum_{t=1}^{T} \bby^{(t)}.
\end{equation}
With this notation in place, we can formally state our problem.

\begin{Problem}\label{P:main}
    Consider an unknown epidemic $\{\bby^{(t)}\}_{t=1}^T$ that spreads over a known graph $\ccalG$. 
    Given the temporally aggregated information $\bbx$ as in~\eqref{E:aggrregation}, estimate the complete evolution of the spread $\{\bby^{(t)}\}_{t=1}^T$.
\end{Problem}

Before presenting our proposed solution to Problem~\ref{P:main} in the next section, a few comments are in order.
First, the observation model is realistic and reasonable in practice.
We get to observe if a person was infected or not over a given period of time as well as the length of the infection (either self-reported or estimated by proxies such as serology tests), but we do not have precise temporal discrimination of when the infection started and ended. 
Second, Problem~\ref{P:main} can be reinterpreted as one of signal demixing in graphs since we are given the (temporal) aggregation of many signals and we want to tell them apart leveraging the graph structure.
This is a challenging and highly underdetermined problem even in the simpler setting of linear network processes.
Finally, to solve this ill-posed and non-linear problem, most existing approaches assume precise knowledge of the epidemic model~\cite{prakash2012spotting, lappas2010finding}.
By constrast, we take a data-driven perspective where we assume nothing about the underlying epidemic model other than it being driven by the topology of $\ccalG$, and we use past known observation pairs $(\bbx, \{\bby^{(t)}\}_{t=1}^T)$ to train our model.
Crucially, only few known pairs are needed and these need not belong to the same graph $\ccalG$ mentioned in Problem~\ref{P:main} since our proposed solution effectively generalizes across graph distributions and sizes, as we illustrate in Section~\ref{sec:results}.

\section{Graph CVAE for temporal reconstruction}
\label{sec:solution}

Assume that we have access to $M$ pairs $(\bbx, \bbY)$ where, for notation simplicity, we have collected the $T$ vectors in $\{\bby^{(t)}\}_{t=1}^T$ as columns of $\bbY \in \reals^{N \times T}$.
Our goal is to leverage these observations to estimate the conditional distribution $p(\bbY|\bbx)$ for the particular scenario of interest.
In this way, given a new observation $\bbx$, we can sample our candidate temporal evolution $\bbY$ from $p(\bbY|\bbx)$.
In determining $p(\bbY|\bbx)$, we adopt a CVAE probabilistic model~\cite{sohn2015learning} with a latent variable $\bbz$ that seeks to model features of the temporal variation of the epidemic that have been collapsed in $\bbx$. 
Intuitively, we want $\bbz$ to capture key aspects of the underlying (unknown) spreading dynamics.
We model $\bbz$ as conditionally Gaussian given $\bbx$, i.e., $p_{\phi}(\bbz|\bbx) = \mathcal{N}(\mu_{\phi}(\bbx), \sigma^2_{\phi}(\bbx))$.
Notice that we have made explicit our focus on a parametric form of the conditional probability, where both the mean and the standard deviation of $\bbz$ are functions of $\bbx$ dependent on the parameters $\phi$.
Following the CVAE framework, we define the distribution of our variable of interest as
\begin{equation}\label{E:distribution_theta}
p_{\theta}(\bbY| \bbx, \bbz) = \ccalN(g_{\theta}(\bbx, \bbz), \sigma_{y}^{2} \bbI).
\end{equation}
Under our probabilistic model, \eqref{E:distribution_theta} reveals that if we have access to the true latent variable $\bbz$ in combination with our observation $\bbx$, then we can apply a parametric \emph{deprojection} function $g_\theta$ to get the expected value of the temporal evolution $\bbY$, where $\sigma_{y}^{2}$ is a common noise variance for all entries. 
From the two conditional probabilities introduced, it follows that we can compute the distribution of interest as
\begin{equation}\label{E:distribution_theta_phi}
    p_{\theta, \phi}(\bbY | \bbx) = \int_{\bbz} p_{\theta}(\bbY | \bbx, \bbz)  p_{\phi}(\bbz|\bbx) \,d \bbz.
\end{equation}
However, solving the integral in~\eqref{E:distribution_theta_phi} can become intractable even for fairly simple parametrizations of $p_{\theta}$ and $p_{\phi}$.
Thus, determining the parameters $\theta$ and $\phi$ that maximize the likelihood of the observed $M$ pairs $(\bbx, \bbY)$ is a challenging endeavor in general.
Instead, we follow the well-accepted route of variational inference and establish a tractable loss inspired by the evidence lower bound (see~\cite{kingma2013auto} for details) that can be optimized via stochastic gradient descent.
To present this loss, we first need to introduce a parametrization for the posterior probability of $\bbz$ given $\bbY$, namely $q_{\psi}(\bbz|\bbY) = \ccalN(\mu_{\psi}(\bbY), \sigma^2_{\psi}(\bbY))$.
We then consider the following compound loss
\begin{align}\label{E:loss}
        \ccalL_{\theta, \phi, \psi}(\bbx, \bbY) = & \, L_1(q_{\psi}(\bbz|\bbY), p_{\phi}(\bbz |\bbx)) + \eta_1 L_2(g_{\theta}(\bbx, \hat{\bbz}), \bbY) \, + \nonumber \\       
        & \eta_2 R_1(\theta, \phi, \psi) + \eta_3 R_2(g_{\theta}(\bbx, \hat{\bbz})),
\end{align}
consisting of two fitting terms $L_1 (\cdot)$ and $L_2(\cdot)$, two regularization terms $R_1 (\cdot)$ and $R_2(\cdot)$, and their relative importance given by the scalar weights $\{\eta_i\}_{i=1}^3$.
In our implementation, we select the loss $L_1 (\cdot)$ as the KL divergence between the argument distributions which ensures that, during testing, our draws from $p_{\phi}(\bbz |\bbx)$ will be close to the more informative draws from $q_{\psi}(\bbz|\bbY)$.
For the reconstruction loss $L_2 (\cdot)$, we recall that the entries of $\bbY$ are binary -- either infected or not -- and select an entry-wise binary cross entropy loss.
In terms of regularization, we implement in $R_1(\cdot)$ an $\ell_2$ penalty on all our parameters.
More interestingly, in $R_2(\cdot)$ we incorporate the knowledge that the evolution of the epidemic should be local in $\ccalG$.
To be precise, we define $R_2(g_{\theta}(\bbx, \hat{\bbz})) = \| \sum_{t=2}^{T} [\hat{\bby}^{(t)} - (\bbA + \bbI) \hat{\bby}^{(t-1)}]_{+} \|_1$, where $[\,\cdot\,]_+$ denotes the positive projection and $\hat{\bby}^{(t)}$ is the $t$-th column of $\hat{\bbY} = g_{\theta}(\bbx, \hat{\bbz})$.
Intuitively, in $R_2(\cdot)$ we penalize the appearance of an infected node at time $t$ when neither itself nor any of its neighbors was infected at time $t-1$.
Finally, notice that the loss in~\eqref{E:loss} is actually a random quantity since the latent variable $\hat{\bbz}$ used in the computation of $L_2(\cdot)$ and $R_2(\cdot)$ is drawn from $q_{\psi}(\bbz|\bbY)$. However, we can still optimize over it via the reparametrization trick~\cite{kingma2013auto}.

We implement the parametric functions in our model -- $g_\theta$ and the mean and standard deviations of $p_\phi$ and $q_\psi$ -- using graph neural networks. 
Our overall deep demixing architecture (DDmix) is illustrated in Fig~\ref{fig:architecture}.
An essential building block of our model is the Graph U-Net (g-U-Net)~\cite{gao2019graph}, which consists of down-sampling and up-sampling graph convolutional network (GCN) layers~\cite{kipf2016semi_supervised}, alongside graph pooling layers (gPool and gUnpool). 
More precisely, the output of a generic GCN layer is given by $\bbH = \mathrm{ReLU}(\tilde{\bbA} \bbH_0 \bbW)$, where $\bbH_0$ is the input to the layer, $\tilde{\bbA}$ is a normalized adjacency matrix, and $\bbW$ is a matrix of trainable weights for that layer.
The gPool layer performs a global max-pooling operation on graph data by introducing a trainable projection vector that maps the features of a given node onto a scalar; see~\cite{gao2019graph} for details. 
On the other hand, the gUnpool layer restores the original graph structure by performing the inverse operation to the corresponding gPool layer. 
In our implementation, the four g-U-Nets included in Fig.~\ref{fig:architecture} have depth equal to one, i.e., they have exactly one pair of pooling and unpooling layers, where the pooling operation reduces in half the number of nodes in the graph.
For each of the g-U-Net blocks, the number of input and output node features (signal values per node) can be seen at the top left and right corners, respectively.

The prior and posterior networks are used to respectively parametrize $p_{\phi}(\bbz|\bbx)$ and $q_{\psi}(\bbz|\bbY)$. 
The prior network takes as input the aggregated graph signal $\bbx$ with a single feature per node and passes it through the g-U-Net block while increasing the number of features per node to $T$. 
Two parallel GCN layers are then applied to the output of g-U-Net, resulting in the mean $\mu_{\phi}$ and standard deviation $\sigma_{\phi}$ of the Gaussian distribution $p_{\phi}$. 
We design the posterior network analogously, with the only difference being that the number of features per node in the input signal $\bbY$ is $T$, corresponding to the complete, uncollapsed time series. 
The obtained parameters are then used to sample from the corresponding distribution (denoted by the red triangles in Fig.~\ref{fig:architecture}), where the posterior network is used during training and the prior network during testing.

Finally, the deprojection network takes as input both the observed $\bbx$ and the drawn latent variable $\hat{\bbz}$ and seeks to output a good estimate $\hat{\bbY} = g_\theta(\bbx, \hat{\bbz})$ of the temporal evolution.
In our implementation, we first pass $\bbx$ through a g-U-Net block while increasing the number of node features to $T$. 
The output of this block is concatenated with $\hat{\bbz}$, resulting in a graph signal with $2T$ features per node. 
This signal is then passed through our last g-U-Net block that combines the information in $\bbx$ and $\hat{\bbz}$ and reduces the number of node features to $T$ to match the length of the time series being estimated.

\section{Numerical experiments}
\label{sec:results}

Through synthetic and real-world graphs, we illustrate the behavior of DDmix in diverse settings.\footnote{Code to replicate the numerical experiments here presented can be found at \href{https://github.com/gojkoc54/Deep_demixing}{https://github.com/gojkoc54/Deep\_demixing}.} In evaluating its performance, we compare DDmix with the following three baselines:

\vspace{1mm}
\noindent i) \textbf{MLP:} 
A multi-layer perceptron (MLP) that takes $\bbx$ as input, propagates it through 3 hidden layers with \{$\frac{1}{4}N T$, $\frac{1}{4}N T$, $N T$\} neurons and ReLU non-linear activations. 
The output layer of size $N T$ represents the vertical concatenation of the columns $\bby^{(t)}$ of $\bbY$.

\vspace{0.5mm}
\noindent ii) \textbf{CNN-nodes:} 
A convolutional neural network (CNN) that takes $\bbx$ as input and performs one-dimensional depthwise convolutions with kernels of size 3 and stride equal to 1. 
No pooling is performed and the number of channels gradually increases from $1$ to $T$ ($1, T/4, T/2, T$) so that the $N\times T$ two-dimensional output represents $\bbY$.

\vspace{0.5mm}
\noindent iii) \textbf{CNN-time:} 
A CNN that performs one-dimensional transposed convolutions~\cite{dumoulin2016guide} over the temporal dimension with fractional strides.
For each node $i$, CNN-time takes $x_i$ as one-dimensional input and after 6 blocks of convolution, batch normalization and elementwise ReLU activation, it computes a $T$-dimensional output representing $[y^{(1)}_i, \ldots, y^{(T)}_i]$.

\vspace{1mm}
\noindent
It should be noted that, although data-driven, the three conventional machine learning baselines considered are graph agnostic. 
More precisely, MLP incorporates fully-connected layers, thus overlooking the notion of locality in $\ccalG$.
CNN-nodes, relying on convolutional filters, assumes a notion of locality inherited by the indexing of the nodes, which need not align with the true notion of locality driving the underlying epidemic.
Finally, CNN-time ignores the effect of interconnections between nodes and seeks to solve the temporal reconstruction for each node independently. 
By contrast, DDmix explicitly incorporates the graph structure in its architecture redounding in higher performance and enhancing its generalizability.


For our synthetic graphs, we use a random geometric graph generator that places $N$ nodes uniformly at random in the unit cube. 
An edge is inserted between two nodes if their Euclidean distance is less than or equal to $d_{r}$. 
For generating graphs of size $N \!\in \!\{100, 250, 500, 1000\}$ nodes, we use parameters $d_{r} \!\in \!\{0.25, 0.15, 0.1, 0.075\}$, respectively.
SIRS parameters (cf. Section~\ref{sec:problem}) are set to $\beta = 0.15$, $\delta = 0.1$ and $\gamma = 0.01$.
We set $\eta_1=\eta_3 = 1$ and $\eta_2 = 10^{-6}$ in the loss~\eqref{E:loss}.
We use the Adam optimizer with a fixed learning rate of $10^{-2}$, $\beta_1$ = 0.9 and $\beta_2$ = 0.999. 
The models are trained in mini-batches of size 4, and a maximum of 50 epochs with auto-stop mechanism in case the validation loss stops dropping for 5 epochs. Unless otherwise stated, we train the models with 4500 realizations of 20-step data on a 100-node random graph and use as figure of merit the mean square error (MSE) of the reconstruction given by $\frac{1}{NT} \|\bbY - \hat{\bbY} \|_\mathrm{F}^2$.

\begin{table}[t]
    \caption{MSE of predictions on unseen graphs with different graph density for two different lengths of the epidemics.  }\label{tab1}
    \vspace{-2mm}
    \centering
    \begin{tabular}{r|cc|cc|cc}
        \hline
        Graph density & \multicolumn{2}{c|}{Baseline} & \multicolumn{2}{c|}{Denser} & \multicolumn{2}{c}{Sparser}\\\hline
        Time steps ($T$)   & 10  & 20   & 10  & 20  & 10  & 20\\\hline
        MLP  & .217   & .233   & .289   & .268  & .254   & .254  \\
        CNN-nodes   & .106  & .188  & .270 & .239   & .177 & .232 \\
        CNN-time   & .143 & .194    & .267  & .240    & .168 & .226  \\
        DDmix  & \textbf{.101}   & \textbf{.177}  & \textbf{.160}   & \textbf{.197}   & \textbf{.075}   & \textbf{.216} \\
        \hline
    \end{tabular}
    \vspace{-2mm}
\end{table}

\vspace{1mm}
\noindent
{\bf Varying the density of the graph.}
In this experiment, we test the candidate models using data randomly generated on three different graphs with the same number of nodes (N=100) as in the training graph but different underlying topology. 
To be specific, their graph densities are respectively the same ($d_{r}'=d_{r}$), denser ($d_{r}'= 1.2 d_{r}$) and sparser ($d_{r}'=0.7 d_{r}$) compared to the baseline density of the training graph. 
For each test graph, we test the temporal reconstruction of  $T=10$ and $T=20$ consecutive steps. 
Table~\ref{tab1} summarizes the models' performance revealing that DDmix significantly outperforms the baseline methods. 
This observation aligns well with our hypothesis: Due to their lack of dependence on the underlying graph structure, non-graph-aware methods are unable to adjust their outputs to topologies unseen during training. 
However, the graph convolution based DDmix incorporates the adjacency matrix in its architecture, thus, it is more robust to topological perturbations. 
Provided identical collapsed signals on differently structured graphs, CNN and MLP based models would always output identical reconstructions, whereas DDmix can adjust its reconstruction to the changing topology.

\begin{figure}[t]
\centering
    \includegraphics[width=\linewidth]{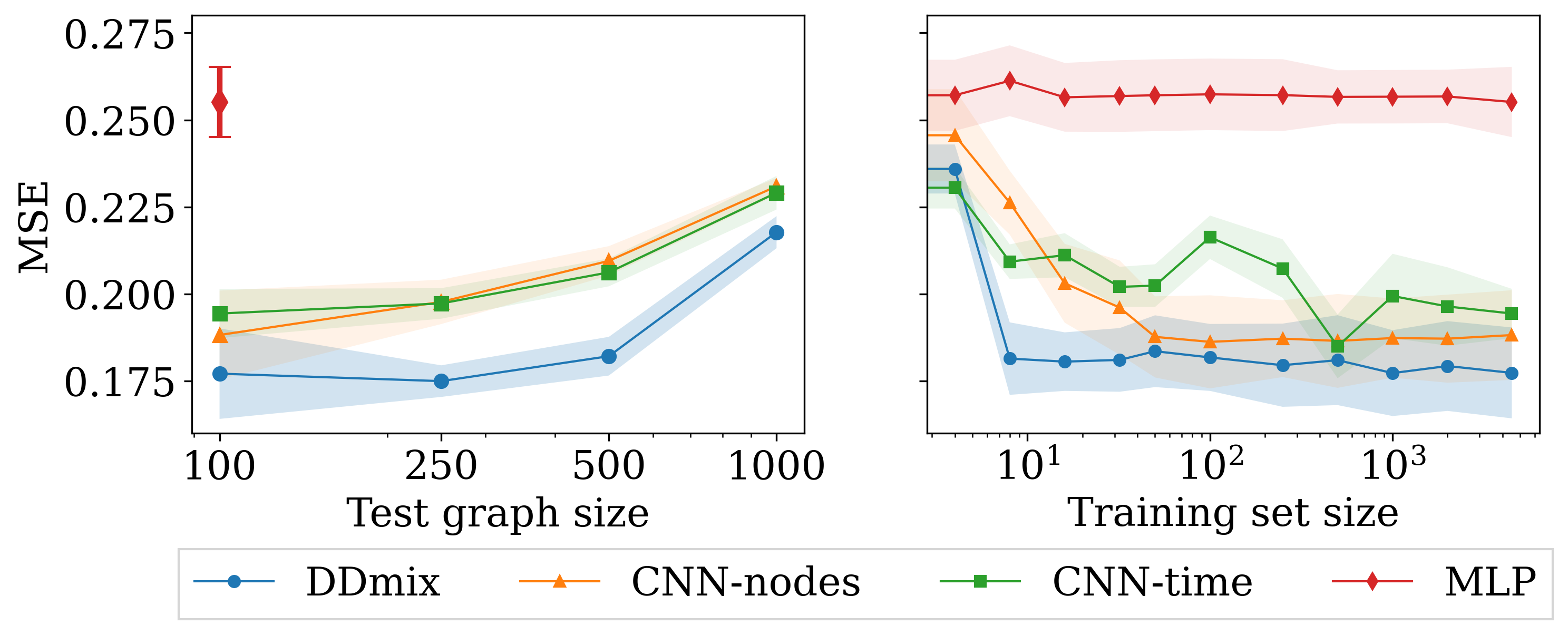}
    \vspace{-5mm}
    \caption{Comparison of models by their temporal reconstruction performance (MSE) against test graph size and training set size. DDmix performs favorably in terms of robustness and generalizability.
    }   
    \vspace{-2mm}
 \label{fig:performance_comparison}    
\end{figure}

\vspace{1mm}
\noindent
{\bf Varying the size of the graph.}
We also tested the models with data generated on graphs of different sizes. 
We fix the number of steps to be $T=20$ and apply models trained on 100-node data to graphs of size $N \in \{100, 250, 500, 1000 \}$. 
As one might expect, the MLP model is unable to handle inputs of varying sizes, thus, can only be tested for $N=100$. 
The variation of MSE for the other methods as a function of test graph size is depicted in Fig.~\ref{fig:performance_comparison} (left). 
The MSE curves in this plot show that DDmix consistently outperforms the competing approaches.
More importantly, DDmix is more robust to changes in the growing size of the test graphs, leading to increased performance gaps for $N \in \{250, 500 \}$.

\vspace{1mm}
\noindent
{\bf Varying the size of the training set.}
Finally, we are interested in the minimal sufficient training set for each model to achieve high accuracy and good generalizability. 
In Fig.~\ref{fig:performance_comparison} (right) we illustrate how MSE generally drops as more data (from 0 to 4500 samples) are used for training each model ($N=100$ and $T=20$).
Initially, all models perform poorly with 4 or fewer training samples. 
Starting at 8 training samples, the models (except for MLP) begin to show improved performance and saturate at around 1000 samples. 
MLP, having the largest number of parameters and not exploiting any locality, cannot effectively learn in this setting.
Most interestingly, DDmix is able to generalize fairly well with only 8 independent training samples, significantly faster than the competing approaches. 
The fact that DDmix learns to recover collapsed signals much easier indicates that our proposed architecture has successfully imposed the right implicit bias by incorporating the graph topology.
\newline

\vspace{1mm}
\noindent
{\bf Reconstructing an epidemic in a primary school.}
We study the identification of epidemic sources in a real-world primary school network. 
Using a public dataset \cite{gemmetto2014mitigation, stehle2011high} containing 2 days of temporal contact among 232 children from 10 school classes, we construct 2 daily contact graphs, one for training ($N\!=\!226$) and the other one for testing ($N\!=\!228$). 
The graphs are undirected and unweighted, and an edge exists between a pair of children if they had more than 5 face-to-face contacts per day. 
Furthermore, during training, a subset of 4 classes is dropped in order to speed up the process and improve generalizability. During testing, the graph has all possible classes and children. 
Synthetic epidemic data contains $T=20$ days of SIRS epidemic evolution with random initial infected locations. 
We simulate 1000 samples for training and 1000 for testing.

Based on this real-world interpersonal network, we evaluate the demixing models by their ability to correctly locate the class where the epidemic started. 
Given a reconstructed $\hat{\bbY}$, we determine a top-$k$ ranking of the source class as follows. First, we determine the first day for which a prediction larger than $0.5$ was made for at least one node, i.e., the smallest $t'$ for which there exists some $i$ such that $\hat{y}^{(t')}_i >0.5$.
We then focus on day $t'$ and rank the classes based on their largest prediction of infection, i.e., we rank each class $\ccalC$ based on $\max_i \hat{y}^{(t')}_i$ for all nodes $i \in \ccalC$.
We say that a top-$k$ ranking yields accuracy equal to $1$ if it contains the true source class and $0$ otherwise.
The average accuracy of DDmix is ${\{25.9, 61.2, 82.9\}}\%$ for the top-1, 3 and 5 predictions, respectively.
This significantly outperforms CNN-nodes' ${\{12.8, 36.8, 60.0\}}\%$ and CNN-time's ${\{11.6, 30.4, 53.5\}}\%$ mean accuracies. 
Accurately tracing the source of epidemics is an increasingly important problem, thus motivating part of our future work.

\section{Conclusions and future work}
\label{sec:conclusions}

We introduced DDmix, a novel graph CVAE architecture for temporal reconstruction of network epidemics from aggregated observations.
Being agnostic to the model of epidemic spread, DDmix relies on the network topology and the observation of past epidemics to solve this inverse (temporal demixing) problem.
Through numerical experiments, we showed that DDmix outperforms non-graph-aware learning techniques in accuracy as well as generalizability across graph distributions and sizes.
Current and future research goals include: i)~The use of DDmix for epidemic prediction and its evaluation on a variety of real epidemic processes, and 
ii)~The development of architectures geared towards the identification of key epidemic actors (such as sources or super-spreaders) without the need for full temporal reconstruction, and iii)~The study of other observation models where data is only observed at a subset of nodes or temporal data is randomly missing.

\vfill\pagebreak


\bibliographystyle{IEEEbib}
\bibliography{strings,refs}

\end{document}